# Joint Optimization of Multimodal Transit Frequency and Shared Autonomous Vehicle Fleet Size with Hybrid Metaheuristic and Nonlinear Programming

**Max T.M. Ng**
*Northwestern University Transportation Center, 600 Foster Street, Evanston, IL 60208, USA*
maxng@u.northwestern.edu

**Hani S. Mahmassani**†
*Northwestern University Transportation Center, 600 Foster Street, Evanston, IL 60208, USA*

**Draco Tong**
*Northwestern University Transportation Center, 600 Foster Street, Evanston, IL 60208, USA*
draco.tong@northwestern.edu

**Ömer Verbas***
*Argonne National Laboratory, Lemont, IL 60439, USA*
omer@anl.gov

**Taner Cokyasar**
*Argonne National Laboratory, Lemont, IL 60439, USA*
tcokyasar@anl.gov

† Deceased * Corresponding author

**Abstract**: Shared autonomous vehicles (SAVs) bring competition to traditional transit services but redesigning multimodal transit network can utilize SAVs as feeders to enhance service efficiency and coverage. This paper presents an optimization framework for the joint multimodal transit frequency and SAV fleet size problem, a variant of the transit network frequency setting problem. The objective is to maximize total transit ridership (including SAV-fed trips and subtracting boarding rejections) across multiple time periods under budget constraints, considering endogenous mode choice (transit, point-to-point SAVs, driving) and route selection, while allowing for strategic route removal by setting frequencies to zero. Due to the problem's non-linear, non-convex nature and the computational challenges of large-scale networks, we develop a hybrid solution approach that combines a metaheuristic approach (particle swarm optimization) with nonlinear programming for local solution refinement. To ensure computational tractability, the framework integrates analytical approximation models for SAV waiting times based on fleet utilization, multimodal network assignment for route choice, and multinomial logit mode choice behavior, bypassing the need for computationally intensive simulations within the main optimization loop. Applied to the Chicago metropolitan area's multimodal network, our method illustrates a 33.3% increase in transit ridership through optimized transit route frequencies and SAV integration, particularly enhancing off-peak service accessibility and strategically reallocating resources.

# 1. INTRODUCTION

## 1.1. Motivation

The integration of Shared Autonomous Mobility Services (SAMS) with traditional transit has emerged as a promising direction for improving system performance and service attractiveness (Ng et al., 2024b). While shared autonomous vehicles (SAVs) offer potential solutions for first-mile-last-mile connectivity and service coverage in low-density areas, they present a fundamental question for transit agencies facing the competition from point-to-point SAMS (ridesharing with SAVs): whether to enhance existing transit services or replace them with SAV-based solutions. The improvement of a mode and a route, or investment in SAV feeders, would divert the budget and riders from or to other services due to competition or transfers. This research addresses this decision-making process by developing a joint planning framework of transit service and SAV feeder fleet.

Previous research efforts investigated the joint optimization of multimodal transit frequency and SAV fleet size (Pinto et al., 2020). However, existing approaches struggle with computational tractability when applied to large-scale networks due to the complex interactions between variables and the non-linear, non-convex problem nature. There is a need for efficient optimization methods that can handle city-scale transit networks while capturing the interdependencies between traditional multimodal transit services and SAMS operations across peak/off-peak hours.

## 1.2. Problem Description

This study focuses on the tactical optimization of transit pattern frequency and SAV feeder fleet size given a set of transit patterns (routes) and origin-destination (O-D) demand. We formulate this as a variant of the transit network frequency setting problem (TNFSP) with four key aspects. First, the primary objective is to maximize transit ridership (including both direct transit usage and transit with first-mile-last-mile SAV feeders[1]) under budget constraints across multiple time periods. Second, apart from transit frequencies (of multiple modes), we also optimize SAV feeder fleet size and consider mode choice across transit (and SAV feeders), point-to-point SAMS, and driving. Third, we allow for the strategic removal of transit routes by setting frequencies to zero, enabling resource reallocation to more efficient services (similar to Pinto et al., 2020). Fourth, we estimate boarding rejections with peak occupancy subtracted by capacity in transit routes and SAV feeders, assuming that transit routes are operated in fixed patterns[2] continuously in a period.

The problem's complexity stems from multiple sources: the combinatorial nature of frequencies across patterns, non-linear relationships between waiting time and vehicle requirements, and the inherent non-convexity introduced by mode and route choice models[3]. Consequently, global optimal solution guarantee is infeasible. Relying on random walks to arrive at a meaningful solution in the high-dimensional non-convex solution space would require excessive computationally heavy iterations (for mode and route choices in particular). To address these challenges, we develop an efficient solution framework that combines a metaheuristic solution with local nonlinear programming (NLP) optimization, strategically exploring the complex non-convex solution space while ensuring local optimality. Specifically, we use particle swarm optimization (PSO) as the metaheuristic approach in this study.

---

[1] SAVs are distinguished from human-driven vehicles in this study primarily by its lower operating cost in the long term (Tirachini and Antoniou, 2020). This makes it economically feasible in more scenarios to serve as first-mile-last-mile SAV feeders integrated into the transit network (Ng et al., 2024a), alongside the competing role of a standalone point-to-point SAMS.

[2] A transit line may consist of multiple of service patterns due to operating strategies such as short-turning and express/local (Verbas and Mahmassani, 2013).

[3] The effects on ridership from frequency and capacity are non-trivial due to variation in arrival time, boarding rejection, and seating availability (Schmöcker et al., 2011; Verbas and Mahmassani, 2015).

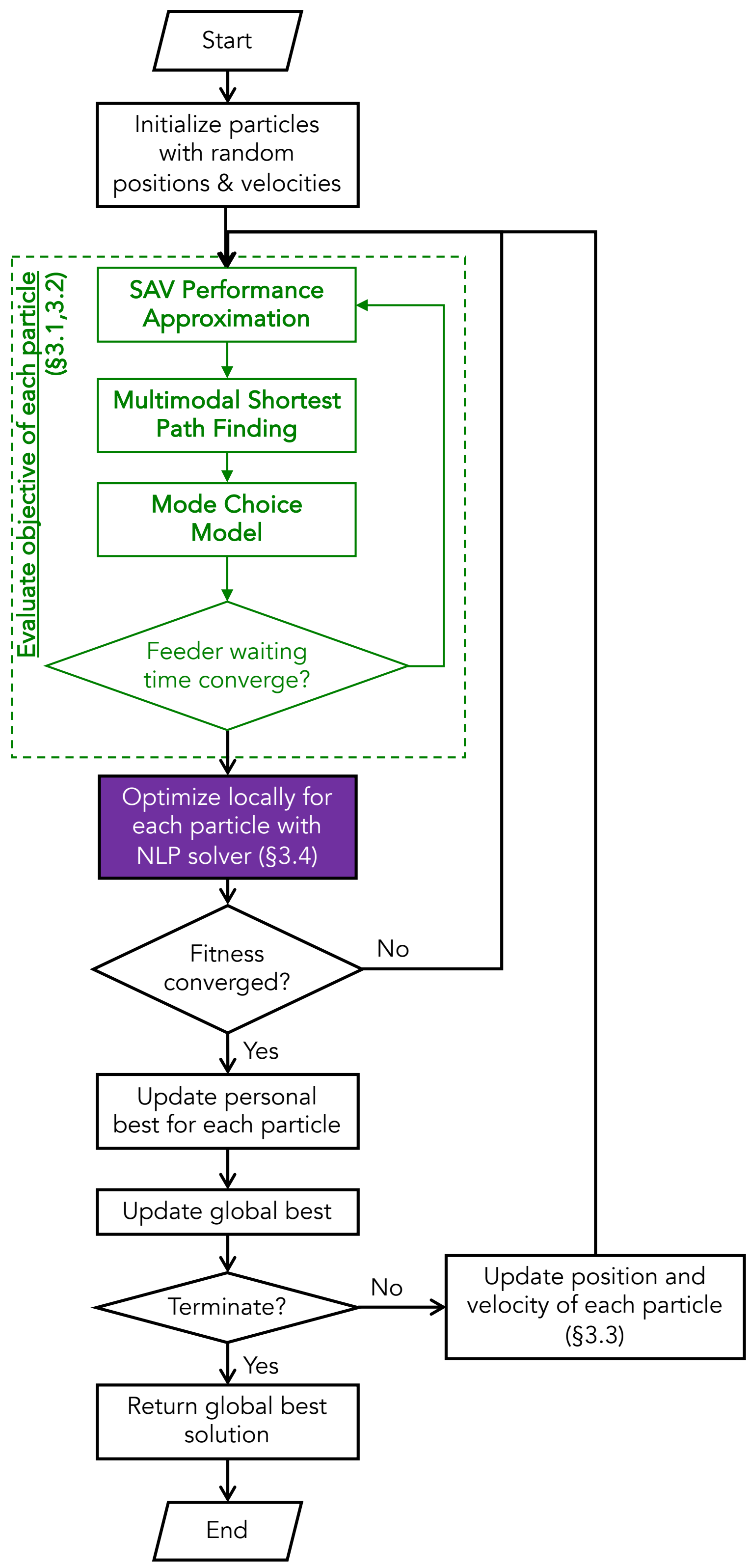

**Figure 1. Flowchart of the solution approach.**

The framework, summarized in Figure 1, incorporates three approximation components to model rider interaction with transit design efficiently without the need for full agent-based simulation: (1) analytical methods for SAMS performance approximation based on fleet size and demand patterns,

(2) a mode choice model using a multinomial logit framework, and (3) traveler assignment approximation with a multimodal transit network with SAV feeders. These models approximate the lower-level problem usually known in bilevel formulations (e.g., Pinto et al., 2020). This integrated approach enables practical application to large-scale urban transit systems while maintaining computational efficiency.

### 1.3. Contributions

This research makes several contributions to transit planning and optimization. Methodologically, we introduce a hybrid optimization framework specifically designed to find high-quality, near-optimal solutions for the large-scale, non-linear, and non-convex multi-period joint transit frequency and SAV fleet size problem in a computationally tractable manner. It combines metaheuristics (PSO) with local NLP optimization, while approximating SAV performance, route choice, and mode choice. The empirical significance is demonstrated through a case study in the Chicago metropolitan area. This real-world application provides insights for transit agencies facing decisions about SAV integration and service planning. The results offer practical guidance for balancing traditional transit services with emerging autonomous vehicle technologies, while considering operational constraints and user preferences.

The remainder of this paper is organized as follows. Section 2 reviews the relevant literature in TNFSP, SAMS, and multimodal transit. Section 3 introduces the formulation, followed by the experiment and results in Section 4 before concluding remarks in Section 5.

## 2. BACKGROUND

### 2.1. Transit Network Frequency Setting Problem (TNFSP)

The TNFSP focuses on frequency optimization and fleet assignment assuming a fixed network design, usually to minimize waiting times for passengers (see reviews by Ceder, 2016; Ibarra-Rojas et al., 2015; Durán-Micco and Vansteenwegen, 2022). Analytical approaches (e.g., Newell, 1971), heuristics (e.g., Furth and Wilson, 1981), and empirical analysis (Ceder, 1984) were earlier preferred solutions to overcome computational limitations. Verbas and Mahmassani (2013) maximized ridership and wait time savings using NLP, later extended to incorporate spatial and temporal demand variations and spatially and empirically-derived elasticities (Verbas et al., 2015a; Verbas and Mahmassani, 2015).

Recent literature has expanded the problem scope. Apart from the typical transit network design and frequency setting problem, features include flow assignment (Martínez et al., 2014), pricing (Bertsimas et al., 2020), schedule setting with bounded stochastic user equilibrium (Jiang et al., 2022), demand uncertainty (An and Lo, 2016), line planning problem (Zhou et al., 2021; Schmidt and Schöbel, 2024), and pattern generation (Ng et al., 2025b). Methodology comprises mixed integer programming (Martínez et al., 2014; Gkiotsalitis and Cats, 2022; Ng et al., 2025b), piecewise linear approximation of waiting time (Zhou et al., 2021), heuristics (Ma and Chow, 2022), and metaheuristics (Aksoy and Mutlu, 2024; Martínez et al., 2014; Gong et al., 2025; see review by Iliopoulou et al., 2019; Chau and Gkiotsalitis, 2025).

1 **Table 1. Comparison of Problem Scopes and Solution Methods of Selected Literature in Transit Network Frequency Setting Problem (TNFSP).**

| Study | Decision Variables | Optimization Objective | Multi-modal | On-demand services | Trip Assignment | Solution Method |
|---|---|---|---|---|---|---|
| Aksoy and Mutlu (2024) | Route frequencies | Minimize total system cost (user, fleet penalty, passenger load penalty) | No | No | Transfer-minimization and logit path choice | Metaheuristics (ABC, DE, FA, GA, PSO) |
| Gkiotsalitis and Cats (2022) | Route frequencies, accommodated passenger demand | Minimize total costs (operational, waiting time, denied boarding) | No | No | Frequency share rule over alternative lines for each origin-destination pair | Mathematical programming (MIQP) |
| Gong et al. (2025) | Route frequencies for different bus classes and time periods | Maximize operating profit or social welfare | No | No | Schedule-based user equilibrium assignment with elastic demand | Metaheuristics (ABC variant) |
| Ma and Chow (2022) | Route frequencies for different time periods | Minimize total costs (operator, user) | Yes | Yes | Multi-agent simulation (MATSim) | Numerical (IOA) |
| Martínez et al. (2014) | Route frequencies from discrete set | Minimize total user travel time (in-vehicle, waiting) | No | No | Frequency share rule | Mathematical programming (MILP); Metaheuristic (TS) |
| Ng et al. (2025) | Route frequencies, stop sequences for different time periods | Minimize total weighted journey time (riding, waiting, transfer time) | No | No | Frequency share rule and passenger flow allocation via MCNF | Mathematical programming (MILP) |
| Pinto et al. (2020) | Route frequencies, SAV fleet size | Minimize total traveler disutility (waiting, boarding rejections) | Yes | Yes | Agent-based simulation | Nonlinear Programming (NLP) |
| Zhou et al. (2021) | Route selection, frequencies | Minimize total costs (operator, user) | No | No | Travel cost minimization in mathematical programming | Mathematical programming (MILP) |
| This study | Route frequencies, SAV fleet size for different time periods | Maximize total served transit ridership (including SAV-fed trips minus boarding rejections) | Yes | Yes | Multimodal shortest path assignment and logit mode choice | Hybrid metaheuristic (PSO) and nonlinear Programming (NLP) |

2 *ABC: Artificial Bee Colony; DE: Differential Evolution; FA: Firefly Algorithm; GA: Genetic Algorithm; IOA: Iterative Optimization Algorithm;*
3 *MCNF: Multi-Commodity Network Flow; MILP: Mixed-Integer Linear Programming; MIQP: Mixed-Integer Quadratic Program;*
4 *NLP: Nonlinear Programming; PSO: Particle Swarm Optimization; TS: Tabu Search*

Table 1 summarizes the problem scopes and solution methods of such previous efforts. This study addresses multimodal TNFSP with SAV feeders in instances too large for exact methods (274 patterns in Chicago over three periods in our experiment). While existing literature widely adopted mathematical programming or metaheuristics, we develop an efficient hybrid solution that combines metaheuristics (PSO) for solution space exploration with NLP for local improvement, built on a previous study (Pinto et al., 2020) that also allows transit frequency to set to minimal for elimination.

### 2.2. Shared Autonomous Mobility Service (SAMS) and Multimodal Transit Network

Pinto et al. (2020) established a framework for joint optimization of transit network design and SAV fleet size. Other works studied SAVs as point-to-point ride-pooling services (Levin et al., 2019; Militão and Tirachini, 2021), semi-on-demand feeders (Ng et al., 2024a), modular transit (Cheng et al., 2024) or replacement of transit lines (Mo et al., 2021; Ng and Mahmassani, 2022). Previous solutions to the joint design problem of frequency and demand-responsive feeders include analytical approaches with queuing network models (Liu and Ouyang, 2021), mathematical programming (Luo et al., 2021), and combined (Shi et al., 2025).

The complexity of SAV operations has led to significant research on fleet management and routing strategies (with seminal works: Alonso-Mora et al., 2017; Hyland and Mahmassani, 2018; see broader reviews by Narayanan et al., 2020). Agent-based simulation tools were used to identify service areas and performance of these SAVs as feeders (Ng et al., 2025a). Recent studies also looked into operational perspective of SAV feeders in multimodal transit network, such as fleet size and vehicle capacity, mostly recommending larger fleets of smaller vehicles (Fielbaum, 2020; Huang et al., 2022; Shan et al., 2024), which also favor more flexible service forms (Ng et al., 2025a, 2024a).

Our work advances the state-of-the-art of joint design beyond a previous solution (Pinto et al., 2020) in three key aspects: (1) direct optimization of ridership through efficient exploration of solution space, (2) integration of mode choice in the local optimization, and (3) computationally efficient approximation methods for large-scale applications. In contrast to studies using continuous approximation (e.g., Liu and Ouyang, 2021; Shi et al., 2025), we focus on real-world tactical application for transit planners to simultaneously design multimodal transit routes with heterogeneous characteristics (e.g., route length, speed, frequency, and ridership) and explicitly consider transfers among transit patterns and with SAV feeders. This positioning addresses the growing need for efficient methodologies to optimize multimodal transit systems in the era of SAVs.

## 3. MODEL

The problem incorporates transit service frequency setting for fixed-route patterns, SAV fleet size determination for first-mile-last-mile connectivity, integration of mode choice and route selection behavior. The objective is to maximize transit ridership while considering mode choice behavior, route selection, and operational constraints across multiple time periods.

Three modes considered are illustrated in Figure 2: (1) transit mode (direct or with first-mile-last-mile SAV feeders), (2) point-to-point SAMS, and (3) driving. For (1) transit mode, the multimodal directed graph $G(f_p, w)$ comprises demand nodes ($o, d \in \mathcal{N}$) and transit nodes, connected by edges $l \in \mathcal{L}$ . Three types of edges represent different transit activities: (1) Access edges (walking/feeder) connecting demand and transit nodes, with access[4] and waiting costs (calculated based on transit frequency $f_p$ and feeder waiting time $w$); (2) Transit edges connecting transit nodes of each pattern with the travel time as cost; and (3) Transfer edges connecting transit nodes of different patterns with transfer penalty and waiting as cost. Additionally, factors are applied to penalize walking, waiting and transfers. In each solution iteration, the graph is used to approximate route choice for transit journey time and graph loading for peak transit load and SAV feeder demand.

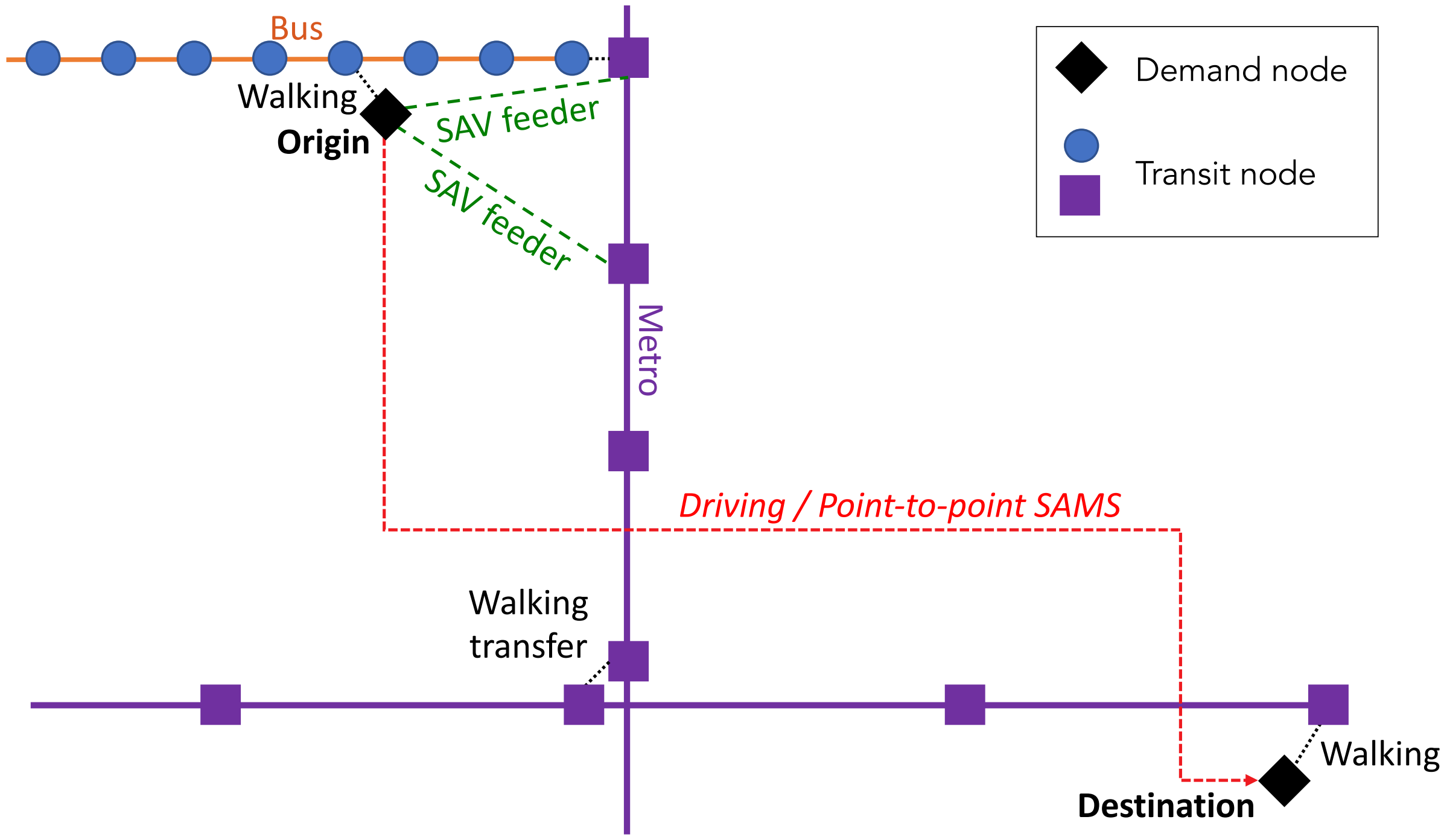

**Figure 2. Illustration of multimodal graphs.**

The overall notation is listed in Table 2. Sets are denoted as capital scripted characters (e.g., $\mathcal{K}$ and $\mathcal{L}_{pk}^{\chi}$), constants as Greek or capital Roman characters (e.g., $\gamma^{\chi}$ and $T_p^{\psi}$), and indices and variables as small Roman characters (e.g., $w_k$ and $f_{pk}$). Superscripts are qualifiers[5] and subscripts are indices. The cardinality of set $\mathcal{P}$ is denoted $|\mathcal{P}|$. All demand, cost, and measurement are expressed per hour unless noted otherwise.

**Table 2. Notation.**

| ***Sets and indices*** | |
|---|---|
| $\mathcal{K}$ | Set of time periods, indexed by $k \in \mathcal{K}$ |
| $\mathcal{N}$ | Set of nodes in network, indexed by $o, d \in \mathcal{N}$ |
| $\mathcal{P}$ | Set of transit patterns, indexed by $p \in \mathcal{P}$ |
| $\mathcal{M}$ | Set of transit modes, indexed by $m \in \mathcal{M}$ |
| $\mathcal{L}$ | Set of links in network, indexed by $l \in \mathcal{L}$ |

[4] The access costs are approximated with constant walking speed for walking edges, and with constant speed, detour, and occupancy for feeder edges. Active mobility (e.g., scooters and bikes) can be considered by updating these parameters. Feeder edges connect demand nodes to transit nodes, e.g., any rail node within a specified radius.

[5] Generally, the following superscripts refer to the three considered modes: τ refers to the transit mode (including trips with SAV feeders), σ refers to the point-to-point SAMS mode, δ refers to the driving mode. Additionally, χ refers to SAV feeder services and links.

| | |
|---|---|
| $\mathcal{L}_{pk}^{\chi}$ | Set of SAV feeder links used to access pattern $p \in \mathcal{P}$ in period $k \in \mathcal{K}$, $\mathcal{L}_{pk}^{\chi} \subset \mathcal{L}$ |
| $\mathcal{L}_{pk}^{\tau}$ | Set of transit links of pattern $p \in \mathcal{P}$ in period $k \in \mathcal{K}$, $\mathcal{L}_{pk}^{\tau} \subset \mathcal{L}$ |
| ***Parameters*** | |
| $A_1, A_2$ | Cut-off utilization in SAV feeder waiting time piecewise linear function |
| $B_m$ | Vehicle size of mode $m \in \mathcal{M}$ |
| $C_{odk}^{\sigma}$ | Driving monetary cost from $o \in \mathcal{N}$ to $d \in \mathcal{N}$ in period $k \in \mathcal{K}$ |
| $C_{odk}^{\delta}$ | Point-to-point SAMS fare from $o \in \mathcal{N}$ to $d \in \mathcal{N}$ in period $k \in \mathcal{K}$ |
| $C_{odk}^{\tau}$ | Transit fare from $o \in \mathcal{N}$ to $d \in \mathcal{N}$ in period $k \in \mathcal{K}$ |
| $D_k$ | Duration of time period $k \in \mathcal{K}$ |
| $F_p^{+}$ | Maximum policy frequency for pattern $p \in \mathcal{P}$ |
| $Q_{odk}$ | Demand from $o \in \mathcal{N}$ to $d \in \mathcal{N}$ in period $k \in \mathcal{K}$ |
| $R^{\sigma}$ | Average SAV feeder occupancy |
| $S_k^{+}$ | Maximum SAV feeder fleet size for period $k \in \mathcal{K}$ |
| $T_p^{\psi}$ | Cycle time of pattern $p \in \mathcal{P}$ (in hour) |
| $T_l^{\chi}$ | Journey time of SAV feeder link $l \in \mathcal{L}_{pk}^{\chi}, p \in \mathcal{P}, k \in \mathcal{K}$ |
| $T_{odk}^{\sigma}$ | Journey time of point-to-point SAMS for traveling from $o \in \mathcal{N}$ to $d \in \mathcal{N}$ in period $k \in \mathcal{K}$ |
| $T_{odk}^{\delta}$ | Journey time of driving for traveling from $o \in \mathcal{N}$ to $d \in \mathcal{N}$ in period $k \in \mathcal{K}$ |
| $U_{odk}^{\sigma}$ | Utility of point-to-point SAMS for traveling from $o \in \mathcal{N}$ to $d \in \mathcal{N}$ in period $k \in \mathcal{K}$ |
| $U_{odk}^{\delta}$ | Utility of driving for traveling from $o \in \mathcal{N}$ to $d \in \mathcal{N}$ in period $k \in \mathcal{K}$ |
| $W$ | Minimum average traveler wait time for SAV feeders |
| $\alpha_1, \alpha_2$ | SAV feeder waiting time fixed-point iterations step sizes for calculation / approximation |
| $\beta_0^{\tau}$ | Mode-specific constant for transit |
| $\beta_1, \beta_2$ | Coefficient for travel time / fare in utility |
| $\delta$ | SAV feeder waiting time fixed-point iterations convergence threshold |
| $\Delta_1, \Delta_2$ | Slopes of SAV feeder waiting time piecewise linear function |
| $\gamma_m^{\tau}$ | Unit operating cost of transit mode $m \in \mathcal{M}$ (in $ per vehicle-hour) |
| $\gamma^{\chi}$ | Unit operating cost of SAMS (in $ per vehicle-hour) |
| $\Gamma^{o}$ | Daily operating budget |
| ***Decision variables*** | |
| $f_{pk}$ | Transit frequency of pattern $p \in \mathcal{P}$ in period $k \in \mathcal{K}$ |
| $s_k$ | SAV fleet size in period $k \in \mathcal{K}$ |
| ***Auxiliary variables*** | |
| $m_p$ | Transit mode used by pattern $p \in \mathcal{P}$ |
| $q_l^{\lambda}$ | Transit demand on link $l \in \mathcal{L}$ |
| $q_{odk}^{\tau}$ | Transit demand from $o \in \mathcal{N}$ to $d \in \mathcal{N}$ in period $k \in \mathcal{K}$ |
| $r_{pk}^{\tau}$ | Transit boarding rejection for pattern $p \in \mathcal{P}$ in period $k \in \mathcal{K}$ |
| $r_{pk}^{\chi}$ | SAV feeder boarding rejection for pattern $p \in \mathcal{P}$ in period $k \in \mathcal{K}$ |
| $u_{odk}^{\tau}$ | Utility of transit (with or without SAV feeders) for traveling from $o \in \mathcal{N}$ to $d \in \mathcal{N}$ in period $k \in \mathcal{K}$ |
| $t_{odk}^{\tau}$ | Transit journey time from $o \in \mathcal{N}$ to $d \in \mathcal{N}$ in period $k \in \mathcal{K}$ |
| $w_k$ | Average SAV feeder waiting time in period $k \in \mathcal{K}$ |
| $\rho_k$ | SAV feeder fleet utilization rate in period $k \in \mathcal{K}$ |

| ***Functions*** | |
|---|---|
| $G(f,u)$ | A multimodal graph based on transit frequency $f$ and SAV feeder waiting time $u$ |
| $TAP(G,q_{od})$ | Traffic assignment problem that returns link flows given multimodal graph $G$ and O-D demand $q_{od}$ |
| $SP(G)$ | Shortest path journey time in a multimodal graph $G$ |
| ***Additional sets in local sub-problem*** | |
| $\mathcal{L}_{odk}^{\phi}$ | Links involved in the shortest path from $o \in \mathcal{N}$ to $d \in \mathcal{N}$ in period $k \in \mathcal{K}$ |
| $\mathcal{P}_{odk}^{\phi}$ | Patterns involved in the shortest path from $o \in \mathcal{N}$ to $d \in \mathcal{N}$ in period $k \in \mathcal{K}$ |
| $\mathcal{R}_{pk}$ | Set of OD pairs that use pattern $p \in \mathcal{P}$ in the shortest paths in period $k \in \mathcal{K}$, $\mathcal{R}_{pk} \subset \mathcal{N} \times \mathcal{N}$ |
| ***Additional parameters in local sub-problem*** | |
| $F_{pk}$ | Reference frequency for pattern $p \in \mathcal{P}$ in period $k \in \mathcal{K}$ |
| $N_{odk}^{\chi}$ | Number of SAV feeder links involved in shortest path rom $o \in \mathcal{N}$ to $d \in \mathcal{N}$ in period $k \in \mathcal{K}$ from approximation models; $N_{odk}^{\chi} \in \{0,1,2\}$ where $N_{odk}^{\chi} = 0$ if no SAV feeder is used; $N_{odk}^{\chi} = 1$ if SAV feeders are used either in access or egress, and $N_{odk}^{\chi} = 2$ if SAV feeders are used in both access and egress |
| $Q_{odk}^{\tau}$ | Reference transit demand from $o \in \mathcal{N}$ to $d \in \mathcal{N}$ in period $k \in \mathcal{K}$ from approximation models |
| $Q_{pk}^{\pi}$ | Reference peak demand from $o \in \mathcal{N}$ to $d \in \mathcal{N}$ in period $k \in \mathcal{K}$ from approximation models |
| $T_{odk}^{\chi}$ | SAV feeder time used in the shortest path from $o \in \mathcal{N}$ to $d \in \mathcal{N}$ in period $k \in \mathcal{K}$ |
| $W_k$ | Reference SAV feeder waiting time in period $k \in \mathcal{K}$ |
| $W^{+}$ | Maximum SAV feeder waiting time |
| $Y^{-}, Y^{+}$ | Lower/upper bound of demand elasticity |
| $\varepsilon$ | Objective convergence threshold |
| $\lambda_i$ | Regularization parameter, $i = 0,1,2$ |
| $\gamma^{\omega}$ | Waiting disutility factor |
| ***Additional auxiliary variables in local sub-problem*** | |
| $q_{pk}^{\phi}$ | Transit demand for pattern $p \in \mathcal{P}$ in period $k \in \mathcal{K}$ |
| $y_{pk}^{\tau}$ | Transit demand elasticity factor for pattern $p \in \mathcal{P}$ in period $k \in \mathcal{K}$ |
| $y_{ik}^{\chi}$ | SAV feeder demand elasticity factor for O-D that uses $i \in \{0,1,2\}$ feeder links in the journey in period $k \in \mathcal{K}$ |
| ***Additional sets in particle swarm optimization*** | |
| $\mathcal{E}$ | Sets of epochs, indexed by $e \in \mathcal{E}$ |
| $\mathcal{I}$ | Sets of particles, indexed by $i \in \mathcal{I}$ |
| ***Additional parameters in particle swarm optimization*** | |
| $P_i$ | Particle swarm optimization hyperparameters, $i = 0,1,2$ |
| ***Additional auxiliary variables in particle swarm optimization*** | |
| $h_i$ | Random variables for particle swarm optimization velocity, $i = 1,2$ |
| $x_{e,i}$ | Solution of particle $i \in \mathcal{I}$ in epoch $e \in \mathcal{E}$ |
| $\widehat{x_{e,i}}$ | Individual best solution achieved by particle $i \in \mathcal{I}$ until epoch $e \in \mathcal{E}$ |
| $x_e^{*}$ | Global best solution until epoch $e \in \mathcal{E}$ |
| $v_{e,i}$ | Velocity of particle $i \in \mathcal{I}$ in epoch $e \in \mathcal{E}$ |

### 3.1. Mathematical Formulation

The overall objective is to maximize served demand (demand $q^{\tau}_{odk}$ minus boarding rejections of transit $r^{\tau}_{pk}$ and SAV feeder $r^{\chi}_{pk}$ of each pattern $p \in \mathcal{P}$)[6] in Eq. (1) across multiple periods $k \in \mathcal{K}$. The decision variables are $f_{pk}$ bounded by policy frequencies $F^{+}_{p}$ in Eq. (2) and SAV feeder fleet size $s_k$ bounded by fleet availability $S^{+}_{k}$ in Eq. (3).

$$\max_{f_{pk},s_k} \sum_{k\in\mathcal{K}} \sum_{o,d\in\mathcal{N}} q^{\tau}_{odk} - \sum_{k\in\mathcal{K}} \sum_{p\in\mathcal{P}} \max\left(r^{\tau}_{pk}, r^{\chi}_{pk}\right) \text{ s.t. (2)-(14)} \tag{1}$$

$$0 \leq f_{pk} \leq F^{+}_{p}, \forall p \in \mathcal{P}, k \in \mathcal{K} \tag{2}$$

$$0 \leq s_k \leq S^{+}_{k}, \forall k \in \mathcal{K} \tag{3}$$

The transit ridership $q^{\tau}_{odk}$ is determined through a multinomial logit mode choice model incorporating utility functions for transit $u^{\tau}_{odk}$, point-to-point SAMS $U^{\sigma}_{odk}$, and driving $U^{\delta}_{odk}$ in Eq. (4). The transit utilities are calculated as the sum of alternative-specific constants (ASCs) $\beta^{\tau}_{0}$, travel time $t^{\tau}_{odk}$ weighted by $\beta_1$, and fare $C^{\tau}_{odk}$ weighted by $\beta_2$ as shown in Eq. (5). Similarly, the utilities for point-to-point SAMS and driving are calculated respectively in Eq. (6)-(7) with travel times $T^{\sigma}_{odk}$ and $T^{\delta}_{odk}$ and monetary cost/fare $C^{\sigma}_{odk}$ and $C^{\delta}_{odk}$.[7]

$$q^{\tau}_{odk} = Q_{odk} \frac{\exp(u^{\tau}_{odk})}{\exp(u^{\tau}_{odk}) + \exp(U^{\sigma}_{odk}) + \exp\left(U^{\delta}_{odk}\right)}, \forall o, d \in \mathcal{N}, k \in \mathcal{K} \tag{4}$$

$$u^{\tau}_{odk} = \beta^{\tau}_{0} + \beta_1 t^{\tau}_{odk} + \beta_2 C^{\tau}_{odk}, \forall o, d \in \mathcal{N}, k \in \mathcal{K} \tag{5}$$

$$U^{\sigma}_{odk} = \beta^{\sigma}_{0} + \beta_1 T^{\sigma}_{odk} + \beta_2 C^{\sigma}_{odk}, \forall o, d \in \mathcal{N}, k \in \mathcal{K} \tag{6}$$

$$U^{\delta}_{odk} = \beta^{\delta}_{0} + \beta_1 T^{\delta}_{odk} + \beta_2 C^{\delta}_{odk}, \forall o, d \in \mathcal{N}, k \in \mathcal{K} \tag{7}$$

The transit journey time $t^{\tau}_{odk}$ is weighted by walking, waiting, and transfers, and based on the shortest path in Eq. (8) in a multimodal graph $G\left(f_{pk}, w_k\right)$, in which the entry arc costs represent the waiting times based on all transit frequency $f_{pk}, \forall p \in \mathcal{P}$ and SAV feeder waiting time $w_k$. The transit and SAV feeder link flows $q^{\lambda}_{l}$ are output from the trip assignment in Eq. (9). The details of both approximations are discussed in Section 3.2.

$$t^{\tau}_{odk} = SP\left(G\left(f_{pk}, w_k\right)\right), \forall o, d \in \mathcal{N}, k \in \mathcal{K} \tag{8}$$

$$q^{\lambda}_{l} = TAP\left(G\left(f_{pk}, w_k\right), q^{\tau}_{odk}\right), \forall l \in \mathcal{L}, k \in \mathcal{K} \tag{9}$$

The SAV feeder waiting time $w_k$ is approximated through a piecewise linear function of utilization $\rho_k$ in Eq. (10) as illustrated in Figure 3 (Pinto et al., 2020). The utilization is calculated with the total SAV feeder time (sum of flows $q_l^\lambda$ and times $T_l^\chi$ of SAV feeder links $\mathcal{L}_{pk}^\chi$) per total vehicle capacity (fleet size $s_k$ multiplied by average occupancy $R^\sigma$) in Eq. (11).[8]

[6] It is assumed that the boarding rejections of transit and SAV feeders would reduce the load for others. For example, insufficient SAV feeder capacity would reduce the transit ridership and therefore the peak occupancy. Therefore, the maximum of two rejection type is taken to calculate the served demand.

[7] The journey times and monetary costs/fares for driving and point-to-point SAMS are treated as fixed exogenous inputs. This assumes that changes in transit service levels do not significantly impact road network congestion.

$$w_k = \begin{cases} W, \rho_k \leq A_1 \\ W + \Delta_1(\rho_k - A_1), A_1 < \rho_k \leq A_2 \\ W + \Delta_1(A_2 - A_1) + \Delta_2(\rho_k - A_2), \rho_k > A_2 \end{cases}, \forall k \in \mathcal{K} \tag{10}$$

$$\rho_k = \frac{\sum_{p \in \mathcal{P}} \sum_{l \in \mathcal{L}_{pk}^\chi} q_l^\lambda T_l^\chi}{R^\sigma s_k}, \forall k \in \mathcal{K} \tag{11}$$

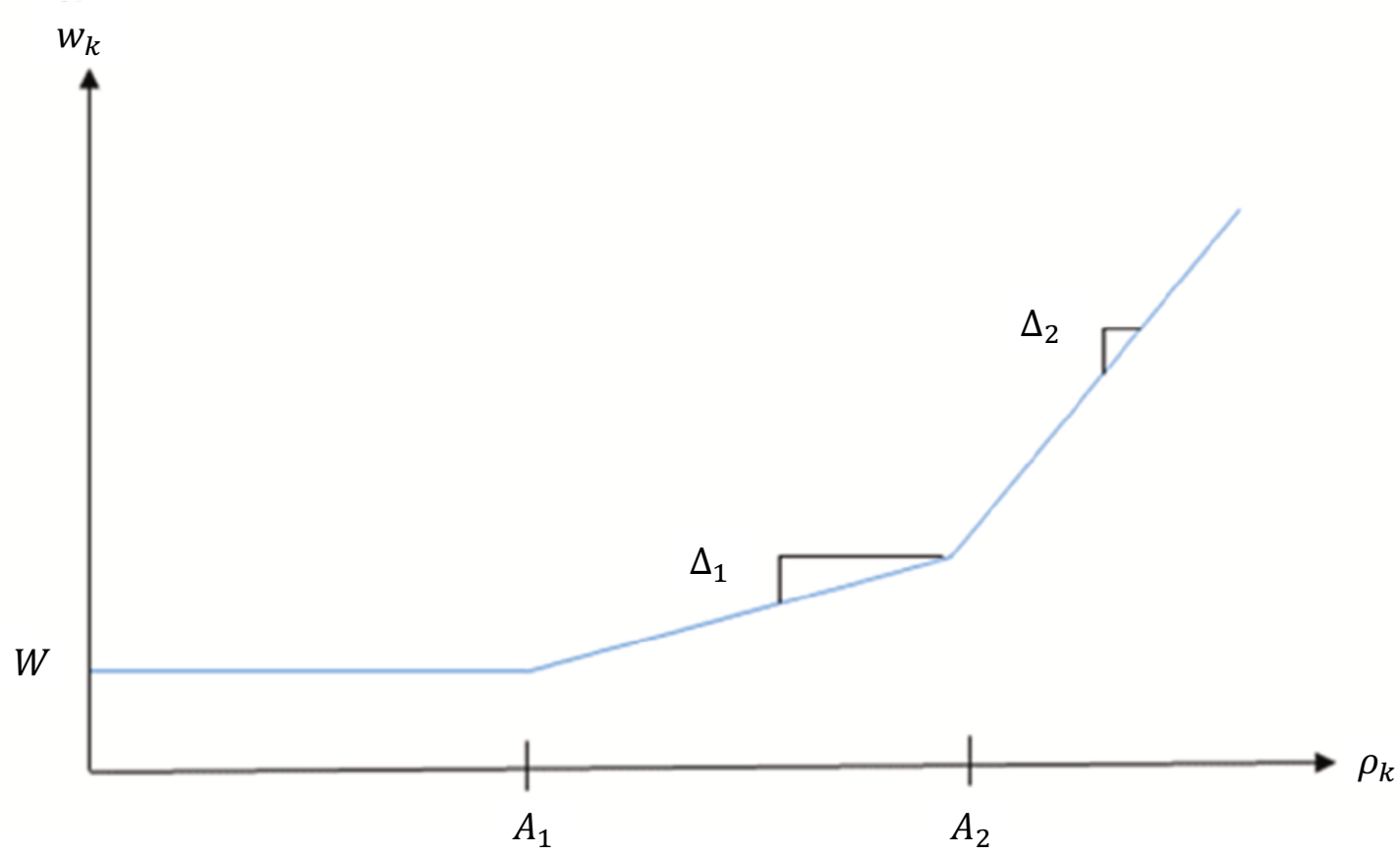

**Figure 3. Piecewise linear relationship between average SAV feeder wait time ($w_k$) and estimated SAV fleet utilization rate ($\rho_k$) (Pinto et al., 2020).**

The boarding rejections are calculated based on the peak transit flow for each pattern and SAV feeders. In Eq. (12), the transit rejection $r_{pk}^\tau$ is the peak flow minus the pattern capacity (vehicle capacity $B_{m_p}$ multiplied by frequency). In Eq. (13), the SAV feeder rejection $r_{pk}^\chi$ is the sum of the unserved portion $(1 - 1/\rho_k)$ of feeder flows.

$$r_{pk}^\tau = \max\left(0, \max_{l \in \mathcal{L}_{pk}^\tau} q_l^\lambda - B_{m_p} f_{pk}\right), \forall p \in \mathcal{P}, k \in \mathcal{K} \tag{12}$$

$$r_{pk}^\chi = \max\left(0, \left(1 - \frac{1}{\rho_k}\right) \sum_{l \in \mathcal{L}_{pk}^\chi} q_l^\lambda\right), \forall p \in \mathcal{P}, k \in \mathcal{K} \tag{13}$$

Finally, the daily budget constraint[9] for all period (each with duration of $D_k$) is set in Eq. (14) by estimating the transit cost with the unit cost $\gamma_{m_p}^\tau$, cycle time $T_p^\psi$, and frequency, and the SAV feeder costs with unit cost $\gamma^\chi$ and the fleet size.

[8] It is possible to expand on the current formulation to optimize for heterogeneous fleet and vehicle sizes of SAV feeders by adjusting $R^\sigma$, $T_l^\chi$, and $\gamma^\chi$ for each service area to reflect the corresponding average occupancy, detours, and operating costs.

[9] Different operating budgeting approaches can be considered with the formulation, e.g., fares or capital costs. The unit operating costs are used here for illustration.

$$\sum_{k\in\mathcal{K}} D_k\left(\sum_{p\in\mathcal{P}} \gamma_{m_p}^{\tau} T_p^{\psi} f_{pk} + \gamma^{\chi} s_k\right) \le \Gamma^{o} \tag{14}$$

### 3.2. Approximation Models

To model rider behavior efficiently without full agent-based simulation, we implement three key approximation components: SAV feeder performance approximation for waiting time in Eq. (10), multimodal graph for route choice in Eq. (8) and (9), and multinomial logit model mode choice in Eq. (4).[10] This subsection discusses the implementation of the multimodal graph and SAV performance approximation in detail.

The multimodal graph in Figure 2 is constructed which consists of nodes representing demand zone centroids and transit stops, and edges representing in-vehicle travel segments, walking links, transfer links, and SAV feeder links, each with an associated cost. The waiting components of graph edge costs are updated in each iteration with new frequency $f_{pk}$ and SAV feeder waiting times $w_k$ for each period $k$. The shortest path is then found for each O-D pair to compute the weighted transit journey time $t_{odk}^{\tau}$ and load the graph with all-or-nothing assignment to calculate the link flow $q_l^{\lambda}$.[11]

For SAV feeder performance approximation, the waiting time $w_k$ for a given demand has no closed-form solution due to its circular dependency with the route and mode choice. To speed up convergence and minimize graph operations required, we employ fixed-point iterations enhanced by analytical approximations:

1. Assume initial SAV feeder waiting time $w_k$ for graph construction.
2. Update total SAV feeder demand $q_l^{\lambda}$ based on trip assignment of the multimodal graph $G(f_{pk}, w_k)$ in Eq. (9).
3. Calculate utilizations $\rho_k$ in Eq. (11) and resulting SAV waiting times $\widetilde{w_k}$ in Eq. (10).
4. Approximate new SAV feeder waiting time $\widehat{w_k}$ with fixed-point iterations using $\widetilde{w_k}$ as starting point and considering demand elasticities, i.e., Eq. (10) and $\widehat{\rho_k} \propto \exp\left(\beta_1 \gamma^{w}(\widehat{w_k} - w_k)\right), \forall k \in \mathcal{K}$.
5. Set new SAV waiting time with a step size between the values from calculation (Step 3) and approximation (Step 4), i.e., $w_k' = \alpha_1 \widehat{w_k} + \alpha_2 \widetilde{w_k}$.
6. Go to Step 2 if $|w_k' - w_k| > \delta$; otherwise end.

### 3.3. Metaheuristics - Particle Swam Optimization (PSO)

The solution methodology combines global exploration through metaheuristics (PSO) (Kennedy and Eberhart, 1995)[12] with local improvement via NLP. This hybrid structure solves the non-convex non-linear optimization problem with the computationally heavy shortest path update and network loading. The approach can be interpreted from two perspectives: as a metaheuristic framework using local NLP optimization as a repair operator, or as NLP optimization with PSO-guided multi-starts.

[10] The framework is compatible with agent-based simulation tools, e.g., NU-TRANS (Verbas et al., 2015b), POLARIS (Auld et al., 2016), and FleetPy (Engelhardt et al., 2022), to replace each approximation model for desired accuracy and allowable computation time.

[11] For simplicity, this study finds the shortest path for each origin to all destinations with Dijkstra's algorithm. Other techniques to handle multimodal graphs such as hyperpath (Verbas et al., 2015b) would allow more accurate modeling of mode and route choices. Smoother traveler assignment instead of all-or-nothing assignment could also help to smoothen the objective function and aid solution finding.

[12] PSO is chosen for its population-based stochastic optimization to explore the non-convex solution space, natural parallelization through independent particle evaluations, and effective representation of solution space through particle positions (frequencies and fleet sizes) and velocities.

The need for global search in this problem arises from several factors that create a highly complex, non-convex solution space:

- The combinatorial nature of frequency settings across hundreds of transit patterns;
- The circular dependency between transit service levels, SAV fleet size, and resulting demand;
- The non-linear relationships in mode choice and waiting time functions; and
- The presence of multiple local optima due to network effects and demand interactions.

The non-convexity hinders assurance that any local minimum found is unique and also the global minimum, and necessitates a global exploration strategy. Besides, the solution space varies significantly depending on the specific problem instance, network structure, and demand patterns. Therefore, the objective of this framework is not to prove global optimality, but rather to develop a computationally efficient method for identifying high-quality, near-optimal solutions for this practical planning problem. The metaheuristics perform the global exploration, and the NLP component (Section 3.4) finds a local optimum within each promising region identified by the PSO.

In this implementation, we utilize PSO as the global search component, though the framework is flexible and can accommodate other metaheuristics. For solution representation, each particle $i \in \mathcal{I}$ of epoch $e \in \mathcal{E}$ represents a complete solution vector $x_{e,i} = [f_{pk}, s_k]$, combining pattern frequencies $f_{pk}$ and SAV fleet sizes $s_k$ for all time periods $k \in \mathcal{K}$.[13]

The particle movement mechanism follows standard PSO dynamics, where each particle's position is updated in Eq. (15) based on last position and velocity $v_{e+1,i}$. In Eq. (16), the velocity is set by three terms: inertia, cognitive (distance to best solution of each particle $\widehat{x_{e,i}}$), and social (distance to global best solution $x_e^*$), with $P_0, P_1, P_2$ as hyperparameters and $h_1, h_2$ as random variables specific to each particle and iteration. These accelerate the particles stochastically towards personal and global best.

$$x_{e+1,i} = x_{e,i} + v_{e+1,i} \tag{15}$$

$$v_{e+1,i} = P_0 v_{e,i} + P_1 h_1 \left(\widehat{x_{e,i}} - x_{e,i}\right) + P_2 h_2 \left(x_e^* - x_{e,i}\right) \tag{16}$$

### 3.4. Local Sub-Problem

The full evaluation of the objective function in Eq. (1) requires running the computationally intensive shortest path and assignment models in Eq. (8)-(9). To refine the solutions provided by the PSO while reducing the PSO iterations needed, we develop a local approximation of the objective function to capture the changes in served demand under new transit frequencies and SAV feeder waiting times. The iteration of this local approximation and updating mode and route choices guides the improvement of each solution with a small step size, so each local optimal solutions are close enough such that route choice and relationship of demand elasticity with respect to waiting times remain unchanged.

[13] To maintain feasibility of each solution with the budget constraint, the remaining budget after providing $s_k$ is used to scale transit frequencies $f_{pk}$.

The local sub-problem keeps the objective of maximizing served demand (transit demand minus rejections), but focuses on modeling the parametrizable components. The modified objective in Eq. (17) maximizes pattern-level demand $q_{pk}^{\phi}$ subtracted by boarding rejections. L2 regularization (with hyperparmeter $\lambda_0$) and local bounds in Eq. (18)-(19) control step size, while improving in the neighborhood of the previous solution $F_{pk}$ and $W_k$.[14]

$$\max_{f_{pk},s_k} \sum_{k\in\mathcal{K}} \sum_{p\in\mathcal{P}} \left( q_{pk}^{\phi} - \max\left(r_{pk}^{\tau}, r_{pk}^{\chi}\right) - \lambda_0\left(f_{pk} - F_{pk}\right)^2 - \lambda_0 (w_k - W_k)^2 \right) \quad \text{s.t. Eq. (3),(10),(14),(18)-(27)} \tag{17}$$

$$0 \le f_{pk} \le \min\left(\lambda_1 F_{pk}, F_p^{+}\right), \forall p \in \mathcal{P}, k \in \mathcal{K} \tag{18}$$

$$\lambda_2 W_k \le w_k \le W^{+}, \forall k \in \mathcal{K} \tag{19}$$

We make use of local elasticity with respect to waiting time to approximate the change in transit ridership. The transit O-D demand $Q_{odk}^{\tau}$ is adjusted by the patterns used for a given pattern $p$ in Eq. (20), where $y_{pk}^{\tau}$ is the elasticity for transit waiting time and $y_{ik}^{\chi}$ is the elasticity for SAV feeder waiting time given the number of SAV feeders used in the journey $i = N_{odk}^{\chi} \in \{0,1,2\}$. These elasticity factors are derived as marginal changes from the multinomial logit model[15] from Eq. (4)-(5), leading to the local approximations in Eq. (21)-(22), whereas $\beta_1$ and $\gamma^{\omega}$ are respectively the utility coefficient of travel time and waiting time factor. See Appendix I for the derivation. Additional bounds are imposed on the factors in Eq. (23) to facilitate numerical stability, as $q_{odk}^{\tau}$ are the product of these factors and may explode with large values.

$$q_{odk}^{\tau} = Q_{odk}^{\tau} \prod_{p|(o,d)\in\mathcal{R}_{pk}} y_{pk}^{\tau}\, y_{ik}^{\chi}\,, i = N_{odk}^{\chi}, \forall o,d \in \mathcal{N}, k \in \mathcal{K} \tag{20}$$

$$y_{pk}^{\tau} = \exp\left(\frac{\beta_1 \gamma^{\omega}}{2}\left(\frac{1}{f_{pk}} - \frac{1}{F_{pk}}\right)\right), \forall p \in \mathcal{P}, k \in \mathcal{K} \tag{21}$$

$$y_{ik}^{\chi} = \exp\left(\beta_1 \gamma^{w} i\, (w_k - W_k)\right), \forall i \in \{0,1,2\}, k \in \mathcal{K} \tag{22}$$

$$Y^{-} \le y_{pk}^{\tau}, y_{ik}^{\chi} \le Y^{+}, \forall p \in \mathcal{P}, k \in \mathcal{K}, i \in \{0,1,2\} \tag{23}$$

We then aggregate O-D level demand $Q_{odk}^{\tau}$ across $(o,d) \in \mathcal{R}_{pk}$ by patterns, incorporating the elasticity factors above in pattern levels in Eq. (24). This is to reduce the number of demand terms in NLP from $O(|\mathcal{N}|^2|\mathcal{K}|)$ to $O(|\mathcal{P}||\mathcal{K}|)$, assuming unchanged route choice and negligible effects of cross-pattern effects on demand.

$$q_{pk}^{\phi} = \sum_{(o,d)\in\mathcal{R}_{pk}} q_{odk}^{\tau} = y_{pk}^{\tau} \sum_{(o,d)\in\mathcal{R}_{pk}|i=N_{odk}^{\chi}} Q_{odk}^{\tau}\, y_{ik}^{\chi}, \forall p \in \mathcal{P}, k \in \mathcal{K} \tag{24}$$

[14] The previous solution ($F_{pk}$ and $W_k$) constant to the local sub-problem are represented by capital letters. The waiting time $w_k$ instead of SAV fleet size $s_k$ is used to simplify the formulation of local sub-problem particularly for the elasticity. They have a one-to-one mapping in Eq. (10)-(11).

[15] The derivation assumes insignificant changes in the logit model denominator for the expected small changes in frequencies and wait times within the local sub-problem. It is likely to hold under lower transit ridership relative to driving and point-to-point SAMS.

SAV waiting time $w_k$ is still calculated with Eq. (10), but the utilization is now approximated in Eq. (25) with adjusted total SAV time $Q^{\tau}_{odk}T^{\chi}_{odk}$ using previous elasticity factors.

$$\rho_k = \frac{1}{R^{\sigma} s_k} \sum_{p \in \mathcal{P}} y^{\tau}_{pk} \sum_{o,d \in \mathcal{R}_{pk} | i = N^{\chi}_{odk}} Q^{\tau}_{odk} T^{\chi}_{odk} y^{\chi}_{ik}, \forall k \in \mathcal{K} \tag{25}$$

In Eq. (26)-(27), transit boarding and SAV feeder boarding rejection are calculated similar to Eq. (12)-(13), except that peak and transit demands ($Q^{\pi}_{pk}, Q^{\tau}_{odk}$) are from the previous solution and adjusted by the elasticity.

$$r^{\tau}_{pk} = \max\left(0, y^{\tau}_{pk} Q^{\pi}_{pk} - B_{m_p} f_{pk}\right), \forall p \in \mathcal{P}, k \in \mathcal{K} \tag{26}$$

$$r^{\chi}_{pk} = \max\left(0, \left(1 - \frac{1}{\rho_k}\right) y^{\tau}_{pk} \sum_{(o,d) \in \mathcal{R}_{pk} | i = N^{\chi}_{odk}} Q^{\tau}_{odk}\, y^{\chi}_{ik}\right), \forall p \in \mathcal{P}, k \in \mathcal{K} \tag{27}$$

Lastly, the budget constraint in Eq. (14) still applies.

This local sub-problem is solved with an NLP solver Knitro using interior-point/barrier direct algorithm. We adopt a multi-start strategy with local perturbations that creates new solutions locally around the previous ones following a distribution. This further enhances exploration of the local solution space. To facilitate numerical stability, we employ two mechanisms. The L2 regularization term in Eq. (17), controlled by the hyperparameter λ, penalizes large deviations from the PSO's provided solution. Concurrently, the local bounds in Eq. (18)-(19) and (23) constrain the solution space, preventing the NLP solver from taking excessively large or infeasible steps and helping to guide it toward a stable convergence.

As shown in Figure 1, the local sub-problem optimization is run after and before the approximation models to evaluate the fitness and update the route and mode choices. The repeated NLP-approximation runs also mean that the solution can move in a small magnitude each step and wait for the approximation to update the route and mode choice, before another iteration of NLP on the new solution. The iteration is re-run until the objective converges, i.e., the change magnitude is smaller than a threshold $\varepsilon$. This ensures sufficient exploitation before passing the solutions to metaheuristics for further exploration.

## 4. EXPERIMENT AND RESULTS

The model is applied to the Chicago metropolitan area transit network to optimize frequencies of 274 transit routes and SAV feeder fleet size in a large-scale urban setting. The multimodal transit network configuration incorporates modes: commuter rail (Metra), metro rail (Chicago Transit Authority, CTA), and bus services (CTA and Pace). The network is constructed with GTFS dataset. Transfer connections are established between routes within walkable distance. Access connections are set from demand zones as walking links to the closest transit modes, and SAV feeder links to rail nodes within 5 km but beyond walking distance.

For the driving and point-to-point SAMS modes, the journey times and costs are estimated based on a road network extracted with OSMnx (Boeing, 2024) with congestion factors. Point-to-point SAMS also include a detour factor $\phi = 1.3$ (Ng et al., 2024b).

Two periods are considered in a day, covering peak and off-peak hours. Demand data are extracted from the activity-based demand model CT-RAMP (Chicago Metropolitan Agency for Planning, 2023). Mode choice model is applied that separates car owners and non-owners. The operating costs are the average U.S. revenue-hourly costs from the National Transit Database (Federal Transit Administration, 2021). The mode choice model and SAV performance approximation parameters are set with reference to Pinto et al. (2020).

Operating budgets are set to align with current transit agency expenditures. The coefficients of walking and waiting are set with reference to Wardman (2004).
A baseline case is used to compare with the optimization results and as the initial solution for warm-starting.[16] The parameter values are summarized in Table 3. The solution vector is of size 550 (274 route frequencies and 1 SAV fleet size for each of 2 periods).

**Table 3. Input parameters of the models.**

| $\lvert\mathcal{K}\rvert$ | 2 |
|---|---|
| $\lvert\mathcal{N}\rvert$ | 41,186 |
| $\lvert\mathcal{P}\rvert$ | 274 |
| $\mathcal{M}$ | [Metra rail, CTA rail, CTA bus, Pace bus] |
| $\lvert\mathcal{L}\rvert$ | 208,375 |
| $\lvert\mathcal{E}\rvert$ | 30 |
| $\lvert\mathcal{I}\rvert$ | 40 |
| $A_1, A_2$ | 0.5, 0.8 |
| $B_m$ | [1200,1000,70,70] passenger (pax)/veh |
| $C^{\tau}_{odk}$ | \$2.5 |
| $C^{\sigma}_{odk}$ | \$(4.85 + 0.2 * time in min + 0.5 * distance in km) |
| $C^{\delta}_{odk}$ | \$(0.54 * distance in km) |
| $F^{+}_{p}$ | 20 service/h |
| $S^{+}_{k}$ | 10,000 AVs |
| $W$ | 3 min |
| $\alpha_1, \alpha_2$ | 0.8, 0.2 |
| $\beta^{\tau}_0, \beta^{\sigma}_0, \beta^{\delta}_0$ | -1.5, -1.7, 0 |
| $\beta_1, \beta_2$ | -0.12/min, -0.5/\$ |
| $\delta$ | 3 min |
| $\Delta_1, \Delta_2$ | 20 min, 50 min |
| $\gamma^{\tau}_{m}$ | \$[4765, 2033, 188, 188]/h |
| $\gamma^{\chi}$ | \$26.4/h |
| $\Gamma^{o}$ | \$10M/day |
| $\varepsilon$ | 0.1 pax |
| $\lambda_0, \lambda_1, \lambda_2$ | 0.001, 2, 0.5 |
| $U^{+}$ | 60 min |
| $Y^{-}, Y^{+}$ | 0, 3 |
| $\gamma^{\omega}$ | 1.5 |
| $P_0, P_1, P_2$ | 0.9, 2.0, 2.0 |

The optimization was performed on a computing cluster with the Red Hat Enterprise Linux 7.9 operating system, Intel Xeon Gold 6338 2GHz, 32 cores, and 224 GB RAM with 8 threads. The solutions are implemented with Pymoo (Blank and Deb, 2020) for metaheuristics and Knitro (Byrd et al., 2006) for the NLP solver. Each PSO+NLP iteration takes around 3 hours.

[16] The peak-hour frequency in the baseline case is 4 for commuter rail, 12 for metro rail, 7 for city bus, 4 for suburban bus; non-peak frequency is 1 for commuter rail, 8 for metro rail, 5 for city bus, 2 for suburban bus.

### 4.1. Computational Performance

The hybrid PSO-NLP method demonstrates superior computational efficiency compared to PSO-only implementation as shown in Figure 4. The first PSO-NLP iteration results show the effectiveness of NLP-based local optimization, even before PSO's global search takes effect. In Figure 4a, the first iteration achieved a 24.8% increase in total served demand compared to the baseline. The PSO-NLP approach achieved rapid convergence to high-quality solutions, while traditional PSO struggled to surpass an objective of 400,000 after deviating from the initial solutions in Figure 4b. Through global exploration of the solution space, later PSO-NLP enhanced the solution at around 14$^{th}$ iteration, demonstrating the value of global search that provides better starting points for NLP to work on.

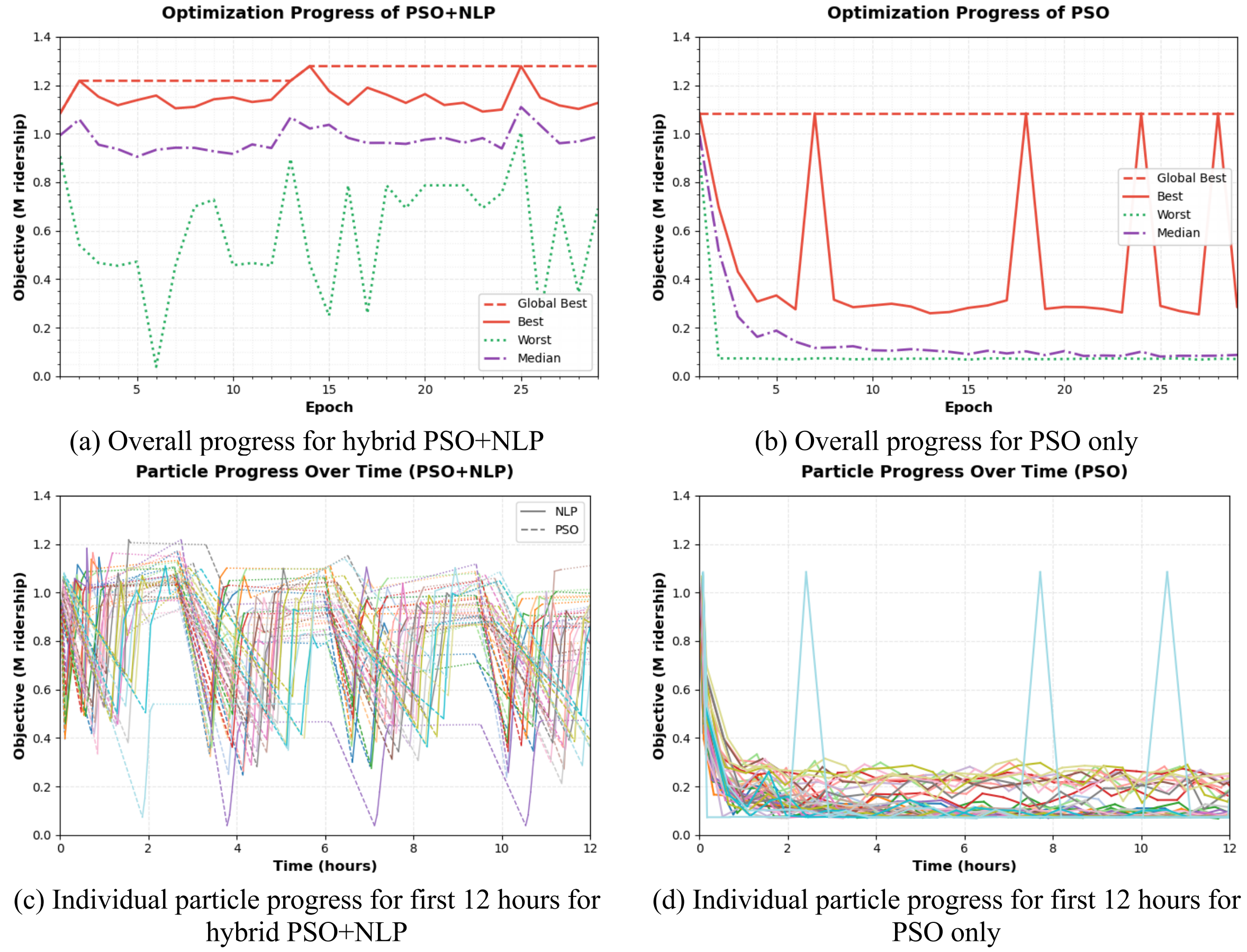

(a) Overall progress for hybrid PSO+NLP

(b) Overall progress for PSO only

(c) Individual particle progress for first 12 hours for hybrid PSO+NLP

(d) Individual particle progress for first 12 hours for PSO only

**Figure 4. Optimization progress.**

The combination of PSO's global exploration capabilities with NLP's local improvement mechanism is highly effective that most solutions maintain high quality after local improvement, albeit with a longer computation time required (around 95 hours compared to 12 hours of PSO alone). The individual particle trajectories (Figure 4c) show frequent local improvements through NLP while maintaining diversity through PSO. This contrasts with Figure 4d where PSO alone fails to improve solutions higher or even to the level of initial solution (except the occasional revisit to the initial solution), likely due to the highly non-convex problem nature that it is unlikely to arrive at a quality solution (reasonable combination of transit frequency and SAV fleet size) randomly.

While the NLP-approximation iterations are computationally intensive, they are essential for solution refinement. The PSO component facilitates broader exploration of the solution space, discovering promising solution regions that were subsequently optimized by NLP. This hybrid approach maintains solution diversity while ensuring effective local optimization, resulting in more robust and higher-quality solutions compared to single-method approaches.

### 4.2. Solution Results

The implementation of the proposed hybrid PSO-NLP optimization framework yields significant improvements in system performance. Table 4 compares the baseline case with both the first iteration best solution (NLP-only) and the global optimal solution (PSO+NLP).

The first iteration results demonstrate the effectiveness of NLP-based local optimization, even before PSO's global search takes effect. Compared to the baseline, the first iteration achieved a 24.8% increase in total served demand while only reducing active routes by 3.3% (from 274 to 265). This shows that NLP alone can significantly improve system efficiency through service reallocation. Peak-hour served demand increased modestly by 3.9%, while off-peak served demand nearly doubled with a 97.2% increase. The service frequency adjustments in the first iteration show a clear trend of service consolidation (though less aggressive than the global best solution). Peak-hour total frequency decreased by 30.9%, with reductions across all modes and most significantly suburban bus. Off-peak services saw similar patterns but with one notable exception: suburban rail frequency increased by 15.8%, indicating the model's early recognition of rail's efficiency for off-peak service. The median route frequencies decreased slightly at peak hours (-11.1%) but increased during off-peak (6.6%), suggesting a more balanced service distribution.

The subsequent global search through PSO+NLP further improved these results. The optimal solution significantly reduced the number of active transit routes by 60.6% from 274 to 108 with a fleet of 5,000 SAVs. This restructuring results in a 33.3% increase in total served demand despite fewer routes used, thanks to flexibility provided by the SAVs. The ridership increase comprises a 12.2% peak-hour and 106.7% off-peak increase. This substantial improvement can be attributed to the strategic deployment of SAVs that help to expand coverage, particularly in off-peak hours where demand is not dense enough for traditional transit modes to achieve economy of scale. As for the services provided in each mode, the redesign reduces the total service frequency at peak hours (32.6%) and off-peak hours (63.5%) to reserve budget for SAV feeder operations. Most frequency cut comes from suburban bus services, while other three modes see slight decrease at peak hours and suburban rail sees off-peak frequency improvement to attract more riders. The restructuring consolidates services and provides high frequency at major corridors, as seen from the higher median route frequency (44.3% and 61.9% increases for metro rail and bus at peak hours). The SAV feeder service contributes to 14.4% and 35.4% of trips at peak and off-peak hours respectively. This again shows that SAV feeder plays a bigger role at off-peak hours when demand is less dense. The feeders are busy at peak hours with 14.4 min average waiting time, and in good service at off-peak with 5.9 min waiting time. Overall, the service reallocation matches high-capacity modes to high-demand corridors while using flexible SAV services for lower-density areas.

Figure 5 illustrates the optimized transit networks at peak and off-peak hours, revealing the prevailing hub-and-spoke network of Chicago. Traditional transit modes maintain strong presence in the urban core and along major corridors within and outside the city, while SAVs are deployed to serve particularly areas outside the city center where traditional transit services are less efficient. The maps demonstrate how the reduction in suburban bus routes is compensated by increased rail and metro bus frequencies in core areas. This spatial redistribution of services appears to better match the demand patterns, while SAV helps to maintain accessibility across the service area.

**Table 4. Result comparison between baseline case and optimal solution.**

| Metrics | Baseline | First iteration best | Percentage change over baseline | Global best | Percentage change over baseline |
|---|---|---|---|---|---|
| **Total** | | | | | |
| **Served demand (pax)** | 930,742 | 1,161,785 | 24.8% | 1,240,992 | 33.3% |
| **Number of active routes** | 274 | 265 | -3.3% | 108 | -60.6% |
| **Peak-hour** | | | | | |
| **Served demand (pax / h)** | 120,377 | 125,122 | 3.9% | 135,020 | 12.2% |
| **Total frequency (service / h)** | 1677 | 1159 | -30.9% | 1131 | -32.6% |
| **Metro rail** | 98 | 67 | -31.4% | 97 | -1.1% |
| **Suburban rail** | 34 | 27 | -20.9% | 32 | -5.0% |
| **Metro bus** | 1011 | 757 | -25.2% | 911 | -9.9% |
| **Suburban bus** | 534 | 309 | -42.2% | 91 | -83.0% |
| **Median route frequency (service / h)** | 4.1 | 3.6 | -11.1% | 11.0 | 170.3% |
| **Metro rail** | 12.2 | 9.6 | -21.4% | 17.7 | 44.3% |
| **Suburban rail** | 3.1 | 2.4 | -20.3% | 2.2 | -28.6% |
| **Metro bus** | 8.2 | 6.8 | -16.3% | 13.2 | 61.9% |
| **Suburban bus** | 4.1 | 2.2 | -46.6% | 4.6 | 13.9% |
| **SAV feeder fleet size** | 0 | 5,000 | N/A | 5,000 | N/A |
| **SAV feeder waiting time (min)** | N/A | 14.2 | N/A | 14.4 | N/A |
| **SAV feeder usage per transit rider** | N/A | 15.4% | N/A | 14.4% | N/A |
| **Non-peak** | | | | | |
| **Served demand (pax / h)** | 18,953 | 37,369 | 97.2% | 39,170 | 106.7% |
| **Total frequency (service / h)** | 833 | 557 | -33.1% | 304 | -63.5% |
| **Metro rail** | 49 | 37 | -23.8% | 20 | -58.8% |
| **Suburban rail** | 11 | 13 | 15.8% | 16 | 46.3% |
| **Metro bus** | 506 | 387 | -23.5% | 255 | -49.6% |
| **Suburban bus** | 267 | 120 | -55.1% | 13 | -95.2% |
| **Median route frequency (service / h)** | 2.0 | 2.2 | 6.6% | 3.0 | 47.6% |
| **Metro rail** | 6.1 | 5.8 | -4.5% | 4.9 | -19.4% |
| **Suburban rail** | 1.0 | 2.2 | 114.8% | 2.8 | 171.1% |
| **Metro bus** | 4.1 | 2.7 | -34.3% | 2.5 | -38.2% |
| **Suburban bus** | 2.0 | 1.5 | -27.6% | 3.4 | 67.8% |
| **SAV feeder fleet size** | 0 | 5,000 | N/A | 5,000 | N/A |
| **SAV feeder waiting time (min)** | N/A | 4.8 | N/A | 5.9 | N/A |
| **SAV feeder usage per transit rider** | N/A | 28.8% | N/A | 35.4% | N/A |

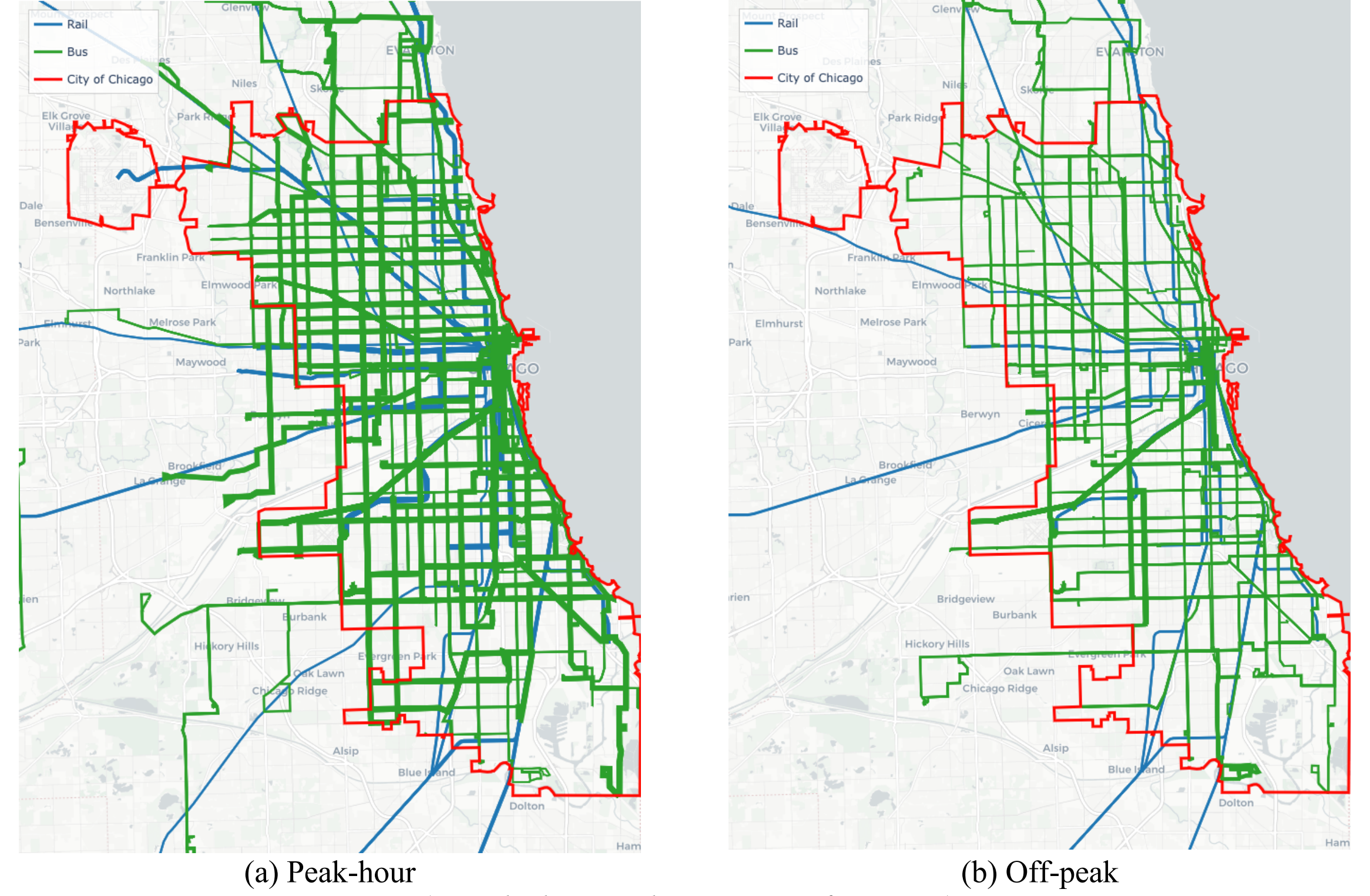

(a) Peak-hour (b) Off-peak

*(Line thickness indicate service frequency)*

**Figure 5. Transit routes in the optimal result.**

# 5. CONCLUSION

## 5.1. Summary

This paper presents a comprehensive framework for jointly optimizing transit service frequency and SAV feeder fleet size in multimodal networks. The proposed methodology addresses several key challenges in transit system planning through: (1) formulation of the joint optimization problem incorporating mode choice and route selection, (2) development of an efficient hybrid solution approach combining metaheuristics (PSO) with NLP in the complex non-convex solution space, and (3) implementation of approximation models for SAV performance and user behavior.

The computational results from the Chicago metropolitan area case study demonstrate the framework's effectiveness in real-world applications. The hybrid PSO-NLP solution approach proves effective, demonstrating the value of combining global exploration with local optimization in addressing complex transit network problems. The approximation models enable efficient evaluation of solutions without compromising the quality of the results. The optimal solution achieves a 33.3% increase in total transit ridership compared to the baseline case, with particularly strong improvements in off-peak hours (106.7%). This improvement is achieved through strategic reallocation of resources, including concentration of traditional transit services in high-demand corridors and strategic deployment of SAV feeders in lower-density areas. SAV feeder integration proved critical to this ridership increase, contributing to 14.4% of peak-hour trips and 35.4% of off-peak trips. This helps to cover low-demand areas with lower operating costs than traditional transit modes.
From a practical perspective, this research provides several insights for transit agencies. While SAV feeder integration can significantly improve system efficiency when strategically deployed, careful consideration must be given to local socioeconomic differences. SAVs facilitate service provision in larger areas with lower costs than current demand-responsive transit. Meanwhile, the current model's uniform treatment of SAV fleet allocation across zones may need refinement to ensure optimal service distribution, particularly in communities where first-mile-last-mile connectivity is crucial. Besides, any changes in essential transit services, particularly involving new service mode like SAV feeders, benefits from community engagement and careful balance of operational efficiency, public service obligations, and community expectations.

## 5.2. Limitations and Future Research

The framework's limitations suggest several directions for future research. First, more sophisticated route choice modeling, e.g., hyperpath, can be integrated in the framework. This would capture a more realistic distribution of passengers across multiple attractive routes, rather than the all-or-nothing assignment used in the current shortest path algorithm. Second, while this paper demonstrates the effectiveness of combining global search with NLP using PSO, future research could explore how different metaheuristics perform within this hybrid framework for various problem instances and network configurations. The choice of metaheuristic may depend on specific problem characteristics such as network size, demand patterns, and computational resources. Third, the effects of transit service and SAV feeder adjustment on road network congestion can be modeled endogeneously. While full Dynamic Traffic Assignment may be computationally prohibitive, macroscopic approaches such as the Network Fundamental Diagram (NFD) could be integrated to approximate the impacts of mode shifts on driving travel times and therefore a more accurate mode shift from driving and point-to-point SAMS. Transit-wise, its effects on road-based transit modes including bus routes can be captured by mapping bus routes to the underlying road network to apply period-specific travel time adjustments. Fourth, the framework can be integrated into comprehensive transit network redesign procedures (Ng et al., 2024b). Fifth, while this study optimizes the fleet size of a homogenous SAV type, heterogeneous design of SAV feeders (including different fleet and vehicle sizes in different service areas) would provide more tailor-made service and better match supply with local demand densities. Lastly, more detailed modeling of mode choice behavior with point-to-point and feeder SAV would better incorporate understanding of user preferences, willingness to transfer, and sensitivity to waiting times across different demographic groups and trip purposes.

These advancements would further strengthen the framework's applicability to real-world transit planning challenges. While the case study is specific to Chicago, the proposed optimization framework is generalizable and can be readily adapted to other metropolitan areas by substituting the corresponding local datasets for the transit network (e.g., GTFS), demand (O-D matrices), and operational costs. Similarly, the model can be applied in more scenarios, e.g., budget increase to determine potential ridership capture, or community-focused service design to ensure optimal distribution of benefits across communities. Additionally, future work could explore fleet pooling between SAV feeders and point-to-point services and the overall integration with Mobility-as-a-Service platforms.

The methodology presented here provides a foundation for transit agencies to make informed decisions about service design and resource allocation in the era of autonomous vehicles. As cities continue to explore innovative transit solutions, this framework offers a practical approach to optimizing multimodal systems while maintaining service quality and operational efficiency.

## ACKNOWLEDGEMENT

The authors would like to dedicate this paper to the memory of our co-author, friend, and mentor, Professor Hani S. Mahmassani, whose guidance was instrumental in the conception and execution of this research.

This material is based on work partially supported by the U.S. Department of Energy, Office of Science, under contract number DE-AC02-06CH11357. This study and the work described were sponsored by the U.S. Department of Energy (DOE) Vehicle Technologies Office (VTO) under the Pathways to Affordable, Convenient and Efficient Regional Mobility, an initiative of the Energy Efficient Mobility Systems (EEMS) Program. Erin Boyd, a DOE Office of Energy Efficiency and Renewable Energy (EERE) manager, played an important role in establishing the project concept, advancing implementation, and providing guidance.

Additional funding was provided by the Northwestern University Transportation Center. The computational experiments were supported in part through the computational resources and staff contributions provided for the Quest high performance computing facility at Northwestern University. The authors remain responsible for all findings and opinions presented in the paper. The contents do not necessarily reflect the views of the sponsoring organizations. Given the simplified assumptions used in the model, this study does not constitute any policy suggestion on current operations.

# APPENDIX - LOCAL SUB-PROBLEM DEMAND ELASTICITY DERIVATION

In Section 3.4, local elasticity factors are used to adjust transit demand with respect to transit pattern frequencies and SAV feeder waiting time. This appendix elaborates on the derivation of these factors in Eq. (21)-(22).

The small step size in the local sub-problem leads to the assumptions of unchanged route choice. Therefore, for each O-D pair, the journey time are only influenced by the waiting time changes in Eq. (A-1). The first term is the change in average transit waiting time (perceived headway). The second term is the change in feeder waiting time, multiplied by the number of feeders used in the journey $N_{odk}^{\chi} \in \{0,1,2\}$. Both are multiplied by waiting disutility coefficient $\gamma^{\omega}$.

$$\Delta t_{odk}^{\tau} = \sum_{p|(o,d)\in\mathcal{R}_{pk}} \frac{\gamma^{\omega}}{2}\left(\frac{1}{f_{pk}} - \frac{1}{F_{pk}}\right) + \gamma^{w} N_{odk}^{\chi}(w_k - W_k), \forall o,d \in \mathcal{N}, k \in \mathcal{K} \tag{A-1}$$

As the journey time is the only changed component of the transit utility function in Eq. (5), the change in utility is calculate in Eq. (A-2) by applying the travel time weight $\beta_1$ in Eq. (A-1).

$$\Delta u_{odk}^{\tau} = \sum_{p|(o,d)\in\mathcal{R}_{pk}} \frac{\beta_1\gamma^{\omega}}{2}\left(\frac{1}{f_{pk}} - \frac{1}{F_{pk}}\right) + \beta_1\gamma^{w} N_{odk}^{\chi}(w_k - W_k), \forall o,d \in \mathcal{N}, k \in \mathcal{K} \tag{A-2}$$

Recall the multinomial logit mode choice model in Eq. (4), we assume insignificant change in the denominator (sum of utilities across three modes) relative to the numerator (transit utility), which is more likely to hold with lower transit ridership relative to driving and point-to-point SAMS. This allows approximating the new transit demand relative to the previous one only based on the change in transit utilities in Eq. (A-3).

$$\frac{q_{odk}^{\tau}}{Q_{odk}^{\tau}} = \frac{\exp(u_{odk}^{\tau})}{\exp(U_{odk}^{\tau})} = \exp(\Delta u_{odk}^{\tau}) \tag{A-3}$$

Combining Eq. (A-2) and (A-3), Eq. (20)-(22) follow.